\documentclass[final,3p,times,twocolumn]{elsarticle}
\usepackage{blindtext, graphicx, amsmath, algorithm, algpseudocode, pifont, comment, layout, amsthm, amssymb}
\usepackage{enumitem}   
\usepackage{eso-pic}
\usepackage{booktabs}
\usepackage{float}
\usepackage{subcaption}

\usepackage{ulem}

\usepackage[utf8]{inputenc}
\usepackage[english]{babel}
\usepackage{hyperref} 
\usepackage{kotex} 
\hypersetup{ colorlinks=true, linkcolor=black, filecolor=black, urlcolor=cyan, }
\usepackage{caption}
\newcommand{\BfPara}[1]{\vspace{0.2em}{\noindent\bf#1.}\xspace}
\journal{ICT Express}
\begin{document}
\begin{frontmatter}
\title{Quantum Distributed Deep Learning Architectures: Models, Discussions, and Applications}
\author{$^{\dag}$Yunseok Kwak}
\ead{rhkrdbstjr0@korea.ac.kr}
\author{$^{\dag}$Won Joon Yun}
\ead{ywjoon95@korea.ac.kr}
\author{$^{\dag}$Jae Pyoung Kim}
\ead{paulkim436@korea.ac.kr}
\author{$^{\ast}$Hyunhee Cho}
\ead{gt7746d@g.skku.edu}
\author{$^{\circ}$Jihong Park\corref{cor1}}
\ead{jihong.park@deakin.edu.au}
\author{$^{\ddag}$Minseok Choi\corref{cor1}}
\ead{choims@khu.ac.kr}
\author{$^{\S}$Soyi Jung\corref{cor1}}
\ead{sjung@hallym.ac.kr}
\author{$^{\dag}$Joongheon Kim\corref{cor1}}
\ead{joongheon@korea.ac.kr}
\address{$^{\dag}$School of Electrical Engineering, Korea University, Seoul, Republic of Korea
\\
$^{\ast}$School of Electronic and Electrical Engineering, Sungkyunkwan University, Suwon, Republic of Korea
\\
$^{\circ}$School of Information Technology, Deakin University, Geelong, VIC, Australia
\\
$^{\ddag}$Department of Electronic Engineering, Kyung Hee University, Yongin, Republic of Korea
\\
$^{\S}$School of Software, Hallym University, Chuncheon, Republic of Korea
}
\cortext[cor1]{Corresponding authors}

\begin{abstract}
Although deep learning (DL) has already become a state-of-the-art technology for various data processing tasks, data security and computational overload problems often arise due to their high data and computational power dependency. 
To solve this problem, quantum deep learning (QDL) and distributed deep learning (DDL) has emerged to complement existing DL methods. 
Furthermore, a quantum distributed deep learning (QDDL) technique that combines and maximizes these advantages is getting attention. This paper compares several model structures for QDDL and discusses their possibilities and limitations to leverage QDDL for some representative application scenarios.
\end{abstract}

\begin{keyword}
Quantum deep learning, distributed deep learning, quantum secure communication
\end{keyword}

\end{frontmatter}

\section{Introduction}
As deep learning (DL) has explosively attracted extensive attention for various data processing tasks, e.g., images, audios, and videos, it has been maturing up to the present \cite{pieee202105park,icdcs2018ahn}.
Basically, the classical DL method requires a large amount of training data, and furthermore, datasets are very complicated and privacy-sensitive especially in some research fields, e.g., finance and medicine. 
In these fields, the classical DL approach is difficult to be applied because of its high demand for computational power and the lack of personal data privacy.
Recently, two methods have been considered as promising methods to deal with these limitations: \textit{Quantum Deep Learning} (QDL)~\cite{wiebe2014quantum} and \textit{Distributed Deep Learning} (DDL). 
Quantum computing utilizes principles in quantum mechanics such as entanglements and superposition, to increase computing efficiency and to save the power consumption.
The most representative QDL techniques that apply quantum computing to DL is \textit{Variational Quantum Circuit (VQC)}~\cite{biamonte2021universal}, which imitates the classical neural  networks with lesser parameters. Also, several studies have shown that QDL can be applied to achieve more efficient and scalable training of classification tasks~\cite{liang2020variational} and deep reinforcement learning (DRL) algorithms~\cite{chen2020variational,YUN20211}.

In parallel, DDL has been studied to solve problems that originate from the centralized setting of storing and computing whole datasets in the classical DL system~\cite{lockwood2021playing}. DDL deals with the computational overhead problem and the data security issue at once. It utilizes the computational power of numerous local devices by operating relatively small models in devices instead of collecting and training all the data at a central server. 
In \textit{Federated Learning} (FL)~\cite{baek2021joint}, local devices train the model with their datasets only and send the locally trained model parameters to the server instead of their datasets.
Meanwhile, \textit{Split Learning} (SL) divides the ANN model, and distributes the front part and the latter one to the local device and the server. Each device then transmits latent variables to the server and the server gives gradients back to the device for training the split model.
Both FL and SL accomplish reducing computational overhead and protecting data security. 
Thus, FL and SL have been applied to such research areas that handle privacy-sensitive data, e.g., medicine~\cite{ha2021spatio,ha2022feasibility,jeon2019privacy}, finance~\cite{long2020federated}, and facial recognition~\cite{kim2021federated}.

In this trend, \textit{Quantum Distributed Deep Learning} (QDDL), which combines the advantages of QDL and DDL, attracts the attention of many researchers. 
QDDL can strengthen data security from external attacks by utilizing secure quantum communication protocols between the server and client. 

Specifically, \textit{Federated Quantum Machine Learning}~\cite{chen2021federated} and \textit{Quantum Federated Learning} (QFL) through \textit{blind quantum computing}~\cite{li2021quantum} utilize QDL algorithms in a federated setting to boost the computational efficiency while ensuring the data privacy and security. In addition, a study~\cite{chehimi2021quantum} proposed the first version of fully-quantum QFL framework by employing quantum federated dataset.

This paper investigates various QDDL techniques and discusses their limitations and challenges. Compared to the conventional QDL and DDL studies, QDDL is still in its infancy and there is no agreements on promising research directions and appropriate architectures. Therefore, this paper broadens this area of study and leverages QDDL for autonomous mobility, a representative application scenario requiring less computations, communication efficiency, and security. 

\begin{figure*}[t!]
    \centering
    \includegraphics[width=0.95\linewidth]{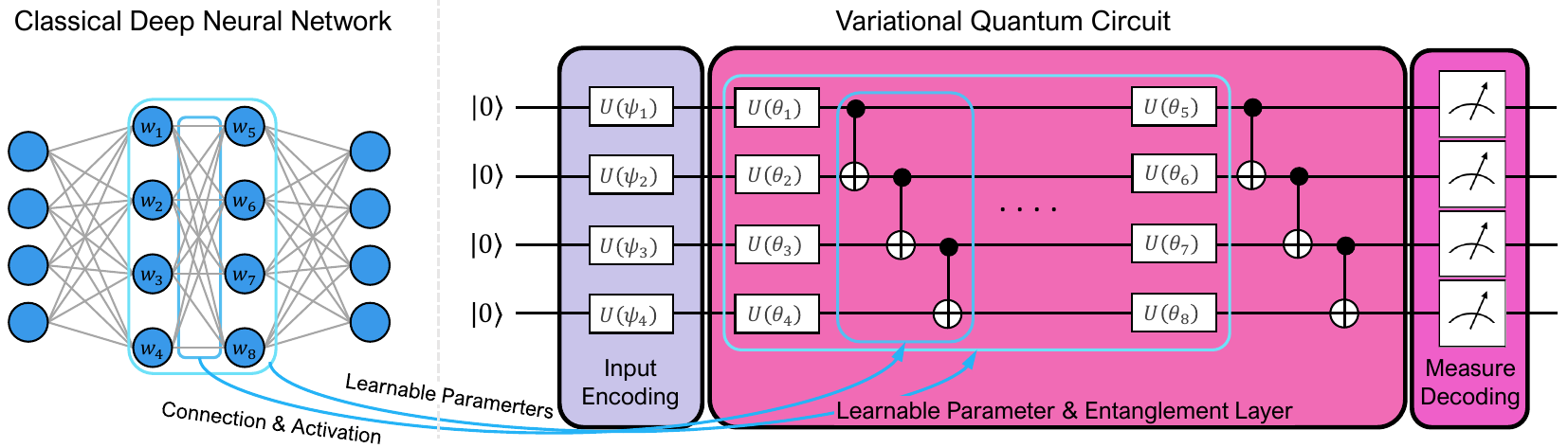}
    \caption{The structure of \textit{Variational Quantum Circuit} (VQC) and classical deep neural network. The parameters and connections with activation function composing classical deep neural network are mapped to the `Rotation' gate and `Controlled Not' gate parameters in the quantum circuit.}
    \label{fig:VQC}
\end{figure*}

\section{Background}

\BfPara{Quantum Deep Learning}
QDL operates numerical calculations of the ANN with the help of quantum computing~\cite{schuld2015introduction}. 
Recently, there have been numerous attempts to replace the classical DL architectures such as \textit{Deep Neural Network} (DNN), \textit{Convolutional Neural Network} (CNN), and \textit{Recurrent Neural Network} (RNN) with quantum-based models as follows. 

\begin{itemize}
    \item \textit{Variational Quantum Circuits:} A variational quantum circuit (VQC) is a quantum circuit that has parameters that can be optimized through the classical DL method~\cite{abohashima2020classification}. Fig.~\ref{fig:VQC} shows the structure of VQC by comparing it with a classical deep neural network. First, training datasets are encoded to have quantum states expressed by qubits. These quantum states are fed into the VQC, which simulate classical neural networks. Then, the output quantum states are measured and collapsed into one of the basic quantum states. 
    Lastly, the output quantum states are converted back into data that we can read and process. VQC can be trained and optimized like any other neural network. VQC, despite its limited size, can be more expressive than classical neural networks while carrying out similar numbers of parameters or with similar learning speeds. To utilize quantum circuits, a quantum encoder is required. Quantum encoders transform classical vectors into quantum states needed as inputs to the quantum circuits. Therefore, quantum encoders play an essential part in the hybrid quantum-classical model because noisy intermediate-scale quantum (NISQ) devices cannot handle large input datasets. Only a limited number of qubits can be used as inputs. Thus, we need highly efficient quantum encoders to express large datasets with few qubits.
    \item \textit{Quantum Convolutional Neural Networks (QCNN):} The QCNN is the quantum circuit with the convolution layer and pooling layer and works in the following five steps~\cite{oh2020qcnn}. First, input data is encoded into its corresponding qubit state and transformed through rotation operator gates. Second, the convolution layer with quasi-local unitary gates filters the input data into a feature map, and the pooling layer with controlled rotation operators downsizes the feature map. Third, this process is repeated, and the fully connected layer acts on the transformed qubit state as classical CNN models. Fourth, the qubit is measured and decoded back into its original classical data form. Finally, circuit parameters are updated by a gradient descent-based optimizer. Through this process, QCNN can process image information as conventional CNN does.
\end{itemize}

\BfPara{Quantum Secure Communication}
The current classical security protocols are particularly vulnerable to attacks utilizing quantum computing. For example, the Rivest-Shamir-Adleman (RSA) public-key cryptosystem, which uses the inability of classical computers to factorize large numbers in a short time to defend against attacks, can be easily defeated by a quantum computer using Shor's algorithm~\cite{shor1994algorithms}. Therefore, the post-quantum cryptography~\cite{bernstein2017post} is necessary to protect against quantum computing-based attacks, and a quantum communication security protocol using quantum entanglement and superposition has been popular for defending against such attacks. 
This paper introduces two well-known quantum secure communication systems, quantum key distribution (QKD)~\cite{scarani2009security} and quantum secure direct communication (QSDC).

\begin{itemize}
    \item \textit{Quantum Key Distribution:} Quantum key distribution is a secure communication method based on quantum mechanical phenomena instead of mathematical complexity, and functions as a way to generate a random secret key. In BB84~\cite{bennett2020quantum}, a representative protocol using QKD, the sender and the receiver use polarization filters to generate and measure photons, and verify the signal in the public channel to use it as a quantum secret key. In communications using QKD, the presence of an eavesdropper is not only known to the sender and the receiver, but also the eavesdropper cannot overhear the accurate information. The security of QKD is based on the characteristic of photons that are non-replicable and collapse when they are measured. 
    
    \item \textit{Quantum Secure Direct Communication:} Quantum secure direct communication (QSDC)~\cite{zhang2017quantum} is another important branch of quantum cryptography. In QSDC, the secret information can be transmitted through a quantum channel directly without sharing a private key. Since QSDC achieves security through coding in quantum terminals, its communication capacity is more efficient than that of QKD because there is no need to manage quantum keys~\cite{long2002theoretically}. 
    At the beginning, the two-step QSDC method has been proposed, which requires two rounds for transmitting photons. This scheme consists of a checking round and a message coding round. In the checking round, the receiver verifies whether the current channel is secure or not, and if the channel security is guaranteed, the receiver directly reads out the encoded message in the message coding round.
    Recently, the one-step protocol~\cite{sheng2022one} has appeared that combines two steps of checking and message coding into one by using polarization-spatial hyperentanglement. This one-step method enables quantum communications at a longer distance compared to existing techniques.
\end{itemize}

\section{Quantum Distributed Deep Learning Architectures}

\begin{figure}[t!]
    \centering
    \includegraphics[width=1.0\linewidth]{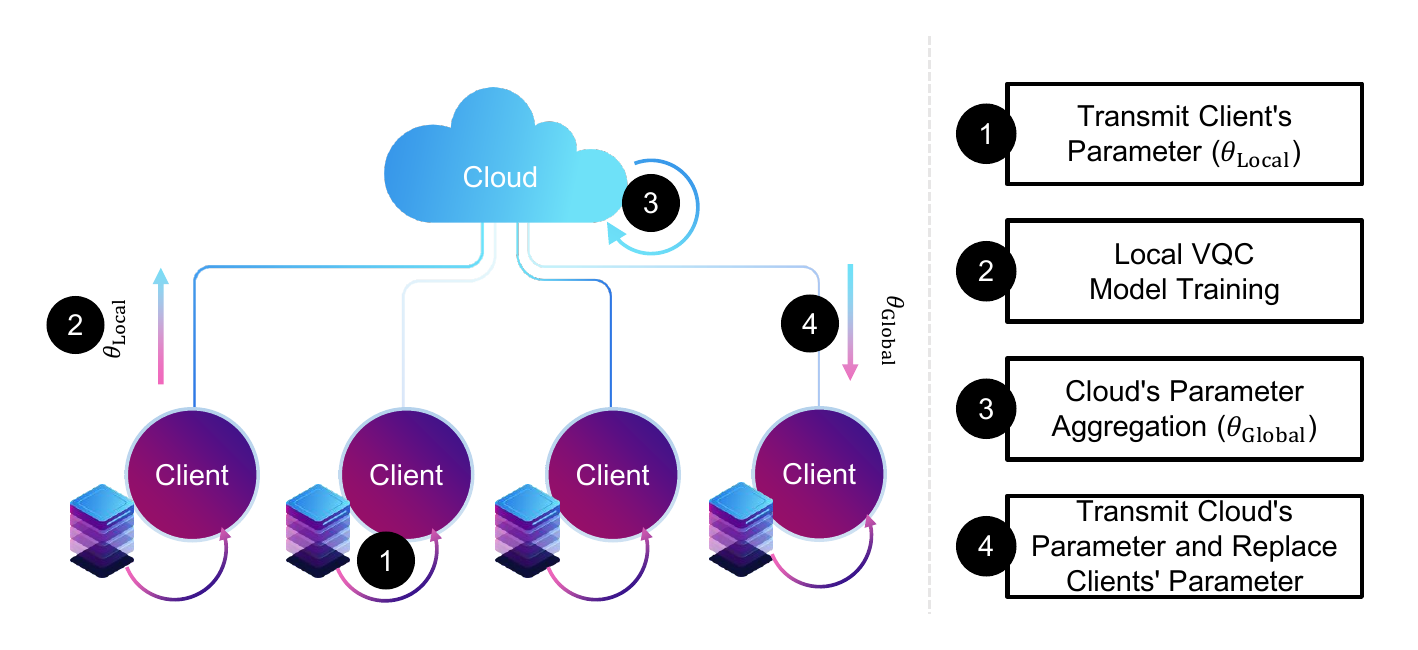}
    \caption{The illustration of QFL.}
    \label{fig:QFL}
\end{figure}

\BfPara{Federated Quantum Machine Learning}
Chen \textit{et al.}~\cite{chen2021federated} proposes the first QML model in the federation setting. Specifically, they considered the VQC and \textit{Quantum Neural Network}s (QNN) coupled with classical pre-trained convolutional neural networks. Fig.~\ref{fig:QFL} shows the illustration of QFL. These hybrid quantum-classical classifiers perform training in a federated setting by using the FedAvg algorithm~\cite{fedavg}. 
This work demonstrated for the first time that QDDL can be implemented in practice by giving following experiments and schematic results.

\begin{itemize}
    \item \textit{Variational Quantum Federated Learning:}
While existing quantum communication security protocols are based on quantum photonic computing, implementation of QDDL is mostly based on ion trapping or superconducting quantum gate operations. therefore, \cite{chen2021federated} selects the VQC instead of quantum communication security protocols to operate the quantum federaed learning algorithm in practical scenarios, and demonstrates its performance for binary classification of the CIFAR image dataset. 
\item \textit{Hybrid Quantum-Classical Transfer Learning:}
The NISQ device is still limited in processing large-sized data due to 
its insufficient computing and error correction capability. Therefore, \cite{chen2021federated} uses and transfers the learning results of a pre-trained classical neural network to learn a quantum nerual network. This method first transforms images into small-dimensional data using a classical CNN, and then processes it in a quantum circuit. Here, a pre-trained VGG16 model~\cite{simonyan2014very} is used for the feature extraction. 

\end{itemize}

\BfPara{QFL through Blind Quantum Computing}
The quantum protocol for distributed learning with remote quantum servers without exposing the private data of local devices has been developed to pursue exponential speedups and privacy protection. Futhermore, according to \cite{li2021quantum}, applying blind quantum computing~\cite{barz2012demonstration} to QFL ensures differential privacy and can safely act under external gradient attacks.

\begin{itemize}
    \item \textit{Blind Quantum Computation:}
Blind quantum computation allows clients to offload calculations of the ANN to a remote quantum server without sharing any sensitive information.
In this protocol, the client is assumed to prepare random qubits chosen from a finite set and send them to the server; therefore, the client does not require any quantum computational power. 
Then, a two-way secure communication channel is set up where the client gives single-qubit measurement instructions to the server. After all the instructions are delivered, the server computes the received clients qubits under the instructions that the client informs, and finally the client receives the desired output from the server. In this process, the server does not receive any data from the client. If there is only a single client, privacy is perfectly protected, and the client can safely utilize the quantum classifier. However, when multiple clients use the identical quantum classifier, attackers can employ the information of the quantum classifier and its updates to retrieve the client's data. 
    \item \textit{Single Client:}
A single-user system can guarantee the perfect security while providing reliable results. The quantum server is perfectly ignorant about the client's quantum classifier from the beginning to the end. Thus,  additional security measures are not needed in a single-user environment.
    \item \textit{Multiple Clients:}
If there are multiple clients, they share the identical quantum classifier model. 
Transmission of parameters, e.g., gradients, to the quantum server for computation could be vulnerable to attacks. If the transmitting parameters are revealed, they can be used for gradient attacks. 
If the updated parameters are leaked, they can be used in a gradient attack. In a gradient attack, the revealed information can be reverse engineered to estimate the inputs of the system to which unauthorized entities are not allowed to have access. To counteract this problem, we use the differential privacy, which adds Gaussian or Laplacian noise to the information being transmitted. 
With the appropriately designed noise mechanism, the attacker cannot extract the critical information from the transmitting signal, while the client still can retrieve the data. 

\item \textit{Gradient Attack:}
If the attacker overhears the transmitted signal from the client to the server, the attacker can use this information to recover the circuit's input.
The attacker can then use this information to recover the circuit's input. The attacker measures the distance between the current gradient and the target to get the information. However, if the differential privacy is considered, the attacker's gradient attack loss function is interrupted because the noise replaces the target gradient. This decreases the accuracy of recovering the input; therefore, the attacker gets the limited information about the input to the system. 

\end{itemize}

\section{Challenges}
\BfPara{Quantum Compatibility Problem}
QML is not stable unlike deep learning and its application has not been used widely yet. It is still limited by the quantum noise without quantum correction when the model exceeds a specific size \cite{preskill2018quantum}. 
A representative method to overcome this limitation is to use a classical ANN as an auxiliary role for quantum computation, as shown in classical-quantum hybrid transfer learning in \cite{chen2021federated}. In addition, most of the quantum secure communication protocols are difficult to be implemented using NISQ devices. 
Therefore, implementation of a complete QDDL architecture including QDL and quantum secure communication protocols on NISQ devices is an important issue.

\BfPara{Data Privacy Problem}
When the sensitive data, e.g., financial and medical data, is handled, it is critical to prevent unauthorized users as well as participants of the training process from having access to other clients' data.
In \cite{sheng2017distributed}, the server and the client use QKD-based quantum communications to exchange information safely. However, the server could still easily identify the client; therefore, this method is not suitable for situations where the security is of paramount importance. On the other hand, in \cite{chen2021federated}, the client and the server did not exchange any data. However, no quantum security protocol is also considered; therefore, this model was especially vulnerable to Byzantine attacks. The model proposed in \cite{li2021quantum} overcomes this challenge by employing the differential privacy technique to block any attacks during data transmission from the client to the server. At the same time, it also proposed a secure communication protocol to prevent the server from identifying which user transmits the data. 
Other possible solutions to the data security issue include SL, first introduced in \cite{gupta2018distributed}. In an SL setting, a part of the entire training network is stored in the client device, and the client and the server exchange latent data and gradients instead of the actual data. As a result, data privacy is protected. This method generally shows faster algorithmic convergence than federated Learning, and application of this SL setting to QNN is expected to ensure the efficient training and data privacy. 

\BfPara{Data Scalability Problem}
It is widely known that the problem of the barren plateau occurs in large-scale QNNs~\cite{mcclean2018barren}, which vanishes the quantum gradient and leads to local convergence. This means that even if the size of the VQC is large enough and many parameters become available, its optimal convergence may become difficult. It is known that this phenomenon is affected not only by the size of the quantum circuit, but also by the quantum calculation error rate, the type of gradient descent method, and the ansatz of the quantum circuit. Therefore, the effective design of quantum circuit ansatz to solve these problems is a big challenge facing whole quantum deep learning research field. To this end, research on efficient quantum encoding~\cite{shee2021qubit} of classical information or research on processing more information inside qubits by constructing a quantum circuit as a tensor network~\cite{qi2021qtn} has been conducted. 

\section{Potential Application for Autonomous Mobility}
The vast amount of data is occurs from mobility platforms (e.g., autonomous driving or advanced aerial mobility) such as visual data (e.g., image, point cloud, scene graph), communication (e.g., wireless communication, quantum communication), and the positional data (e.g., IMU data, GPS data, SLAM). To achieve automated mobility, \textit{cognition, communication, and control} should be considered as follows. 

    \BfPara{Cognition} For an autonomous mobility platform, a machine should be cognitive to collected data. QCNN has shown successful operation in the image classification task, where the image size is small (e.g., MNIST, FashionMNIST, CIFAR10)~\cite{ictc20qcnn}. The basis of cognition around circumstances is object detection. Even the author of~\cite{li2020hierarchical} has suggested object detection using QAOA, the object detection of QDL version is not discussed yet. The VQC-based object detection is an expected topic and it enables QDDL.

    \BfPara{Communication} Building a low-latency secure communication network is essential in creating an autonomous mobile platform. Quantum communication is expected to play a vital role in making such future networks, and research on terrestrial~\cite{inagaki2013entanglement}, drone~\cite{liu2021optical} and satellite-based~\cite{yin2017satellite} quantum networks has been extensively conducted. Current satellite-drone-terrestrial integrated quantum network research uses a method of transmitting encrypted quantum information using a QKD-based security protocol~\cite{hill2017drone}. Combining such an integrated quantum network and QDDL technology is expected to play a decisive role in autonomous mobile-based distributed learning.
    
    \BfPara{Control} Quantum reinforcement learning (QRL) has been actively studied. \cite{kwak2021introduction} have proposed the hybrid computing methods, i.e., the controller policy is based on VQC, and the evaluator-side network (i.e., critic) is based on a classical neural network. In addition, \cite{jerbi2021quantum} have proposed an utterly quantum version of the reinforcement learning regime. Since there are many simulation APIs, e.g., Airsim \cite{airsim} for aerial mobility CARLA \cite{carla} for autonomous driving, the implementation of QRL is expected technology soon. In addition, based on the studies mentioned above, quantum multi-agent distributed reinforcement learning, where the basis is QDDL, is an expected key solution for autonomous mobility platforms. 


\section{Conclusion}
In this work, we first introduced the QDDL studies and their substructures. We also discussed the pros and cons of these studies from multiple perspectives, with a particular focus on data security. We also explored the fields of applications for the research domain and the possibilities of new methodologies. We believe this contribution will definitely help conduct future quantum distributed AI research.


\bibliographystyle{elsarticle-num}

\begin{thebibliography}{10}
\expandafter\ifx\csname url\endcsname\relax
  \def\url#1{\texttt{#1}}\fi
\expandafter\ifx\csname urlprefix\endcsname\relax\def\urlprefix{URL }\fi
\expandafter\ifx\csname href\endcsname\relax
  \def\href#1#2{#2} \def\path#1{#1}\fi

\bibitem{pieee202105park}
J.~Park, S.~Samarakoon, A.~Elgabli, J.~Kim, M.~Bennis, S.-L. Kim, M.~Debbah,
  Communication-efficient and distributed learning over wireless networks:
  Principles and applications, Proceedings of the IEEE 109~(5) (2021) 796--819.

\bibitem{icdcs2018ahn}
S.~Ahn, J.~Kim, E.~Lim, W.~Choi, A.~Mohaisen, S.~Kang, {ShmCaffe}: A
  distributed deep learning platform with shared memory buffer for {HPC}
  architecture, in: ICDCS, 2018.

\bibitem{wiebe2014quantum}
N.~Wiebe, A.~Kapoor, K.~M. Svore, Quantum deep learning, CoRR abs:1412.3489.

\bibitem{biamonte2021universal}
J.~Biamonte, Universal variational quantum computation, Physical Review A
  103~(3) (2021) L030401.

\bibitem{liang2020variational}
J.-M. Liang, S.-Q. Shen, M.~Li, L.~Li, Variational quantum algorithms for
  dimensionality reduction and classification, Physical Review A 101~(3) (2020)
  032323.

\bibitem{chen2020variational}
S.~Y.-C. Chen, C.-H.~H. Yang, J.~Qi, P.-Y. Chen, X.~Ma, H.-S. Goan, Variational
  quantum circuits for deep reinforcement learning, IEEE Access 8 (2020)
  141007--141024.

\bibitem{YUN20211}
W.~J. Yun, S.~Jung, J.~Kim, J.-H. Kim, Distributed deep reinforcement learning
  for autonomous aerial {eVTOL} mobility in drone taxi applications, ICT
  Express 7~(1) (2021) 1--4.

\bibitem{lockwood2021playing}
O.~Lockwood, M.~Si, Playing atari with hybrid quantum-classical reinforcement
  learning, CoRR abs:2107.04114.

\bibitem{baek2021joint}
H.~Baek, W.~J. Yun, Y.~Kwak, S.~Jung, M.~Ji, M.~Bennis, J.~Park, J.~Kim, Joint
  superposition coding and training for federated learning over multi-width
  neural networks, CoRR abs/2112.02543.

\bibitem{ha2021spatio}
Y.~J. Ha, M.~Yoo, G.~Lee, S.~Jung, S.~W. Choi, J.~Kim, S.~Yoo, Spatio-temporal
  split learning for privacy-preserving medical platforms: Case studies with
  {COVID-19 CT}, {X}-ray, and {Cholesterol} data, IEEE Access 9 (2021)
  121046--121059.

\bibitem{ha2022feasibility}
Y.~J. Ha, G.~Lee, M.~Yoo, S.~Jung, S.~Yoo, J.~Kim, Feasibility study of
  multi-site split learning for privacy-preserving medical systems under data
  imbalance constraints in {COVID-19}, {X}-ray, and cholesterol dataset,
  Scientific Reports 12~(1) (2022) 1--11.

\bibitem{jeon2019privacy}
J.~Jeon, J.~Kim, J.~Kim, K.~Kim, A.~Mohaisen, J.-K. Kim, Privacy-preserving
  deep learning computation for geo-distributed medical big-data platforms, in:
  DSN, IEEE, 2019.

\bibitem{long2020federated}
G.~Long, Y.~Tan, J.~Jiang, C.~Zhang, Federated learning for open banking, in:
  Federated learning, Springer, 2020, pp. 240--254.

\bibitem{kim2021federated}
J.~Kim, T.~Park, H.~Kim, S.~Kim, Federated learning for face recognition, in:
  2021 IEEE International Conference on Consumer Electronics (ICCE), IEEE,
  2021, pp. 1--2.

\bibitem{chen2021federated}
S.~Y.-C. Chen, S.~Yoo, Federated quantum machine learning, Entropy 23~(4)
  (2021) 460.

\bibitem{li2021quantum}
W.~Li, S.~Lu, D.-L. Deng, Quantum federated learning through blind quantum
  computing, Science China Physics, Mechanics \& Astronomy 64~(10) (2021) 1--8.

\bibitem{chehimi2021quantum}
M.~Chehimi, W.~Saad, Quantum federated learning with quantum data, CoRR
  abs:2106.00005.

\bibitem{schuld2015introduction}
M.~Schuld, I.~Sinayskiy, F.~Petruccione, An introduction to quantum machine
  learning, Contemporary Physics 56~(2) (2015) 172--185.

\bibitem{abohashima2020classification}
Z.~Abohashima, M.~Elhosen, E.~H. Houssein, W.~M. Mohamed, Classification with
  quantum machine learning: A survey, CoRR abs:2006.12270.

\bibitem{oh2020qcnn}
S.~Oh, J.~Choi, J.~Kim, A tutorial on quantum convolutional neural networks
  (qcnn), in: ICTC, IEEE, 2020, pp. 236--239.

\bibitem{shor1994algorithms}
P.~W. Shor, Algorithms for quantum computation: discrete logarithms and
  factoring, in: FOCS, IEEE, 1994, pp. 124--134.

\bibitem{bernstein2017post}
D.~J. Bernstein, T.~Lange, Post-quantum cryptography, Nature 549~(7671) (2017)
  188--194.

\bibitem{scarani2009security}
V.~Scarani, H.~Bechmann-Pasquinucci, N.~J. Cerf, M.~Du{\v{s}}ek,
  N.~L{\"u}tkenhaus, M.~Peev, The security of practical quantum key
  distribution, Reviews of Modern Physics 81~(3) (2009) 1301.

\bibitem{bennett2020quantum}
C.~H. Bennett, G.~Brassard, Quantum cryptography: Public key distribution and
  coin tossing, arXiv preprint arXiv:2003.06557.

\bibitem{zhang2017quantum}
W.~Zhang, D.-S. Ding, Y.-B. Sheng, L.~Zhou, B.-S. Shi, G.-C. Guo, Quantum
  secure direct communication with quantum memory, Physical Review Letters
  118~(22) (2017) 220501.

\bibitem{long2002theoretically}
G.-L. Long, X.-S. Liu, Theoretically efficient high-capacity
  quantum-key-distribution scheme, Physical Review A 65~(3) (2002) 032302.

\bibitem{sheng2022one}
Y.-B. Sheng, L.~Zhou, G.-L. Long, One-step quantum secure direct communication,
  Science Bulletin 67~(4) (2022) 367--374.

\bibitem{fedavg}
B.~McMahan, E.~Moore, D.~Ramage, S.~Hampson, B.~A. y~Arcas,
  Communication-efficient learning of deep networks from decentralized data,
  in: AISTATS, Vol.~54, 2017, pp. 1273--1282.

\bibitem{simonyan2014very}
K.~Simonyan, A.~Zisserman, Very deep convolutional networks for large-scale
  image recognition, CoRR abs:1409.1556.

\bibitem{barz2012demonstration}
S.~Barz, E.~Kashefi, A.~Broadbent, J.~F. Fitzsimons, A.~Zeilinger, P.~Walther,
  Demonstration of blind quantum computing, Science 335~(6066) (2012) 303--308.

\bibitem{preskill2018quantum}
J.~Preskill, Quantum computing in the {NISQ} era and beyond, Quantum 2 (2018)
  79.

\bibitem{sheng2017distributed}
Y.-B. Sheng, L.~Zhou, Distributed secure quantum machine learning, Science
  Bulletin 62~(14) (2017) 1025--1029.

\bibitem{gupta2018distributed}
O.~Gupta, R.~Raskar, Distributed learning of deep neural network over multiple
  agents, JNCA 116 (2018) 1--8.

\bibitem{mcclean2018barren}
J.~R. McClean, S.~Boixo, V.~N. Smelyanskiy, R.~Babbush, H.~Neven, Barren
  plateaus in quantum neural network training landscapes, Nature Communications
  9~(1) (2018) 1--6.

\bibitem{shee2021qubit}
Y.~Shee, P.-K. Tsai, C.-L. Hong, H.-C. Cheng, H.-S. Goan, A qubit-efficient
  encoding scheme for quantum simulations of electronic structure, CoRR
  abs:2110.04112.

\bibitem{qi2021qtn}
J.~Qi, C.-H.~H. Yang, P.-Y. Chen, Qtn-vqc: An end-to-end learning framework for
  quantum neural networks, CoRR abs:2110.03861.

\bibitem{ictc20qcnn}
S.~Oh, J.~Choi, J.~Kim, A tutorial on quantum convolutional neural networks
  {(QCNN)}, in: ICTC, 2020, pp. 236--239.

\bibitem{li2020hierarchical}
J.~Li, M.~Alam, A.~A. Saki, S.~Ghosh, Hierarchical improvement of quantum
  approximate optimization algorithm for object detection, in: ISQED, IEEE,
  2020, pp. 335--340.

\bibitem{inagaki2013entanglement}
T.~Inagaki, N.~Matsuda, O.~Tadanaga, M.~Asobe, H.~Takesue, Entanglement
  distribution over 300 km of fiber, Optics express 21~(20) (2013)
  23241--23249.

\bibitem{liu2021optical}
H.-Y. Liu, S.-N. Zhu, Optical-relayed entanglement distribution using drones as
  mobile nodes, Physical Review Letters 126~(2) (2021) 020503.

\bibitem{yin2017satellite}
J.~Yin, Y.~Cao, Y.-H. Li, S.-K. Liao, L.~Zhang, J.-G. Ren, W.-Q. Cai, W.-Y.
  Liu, B.~Li, H.~Dai, et~al., Satellite-based entanglement distribution over
  1200 kilometers, Science 356~(6343) (2017) 1140--1144.

\bibitem{hill2017drone}
A.~D. Hill, J.~Chapman, K.~Herndon, C.~Chopp, D.~J. Gauthier, P.~Kwiat,
  Drone-based quantum key distribution, Urbana 51 (2017) 61801--63003.

\bibitem{kwak2021introduction}
Y.~Kwak, W.~J. Yun, S.~Jung, J.-K. Kim, J.~Kim, Introduction to quantum
  reinforcement learning: Theory and pennylane-based implementation, in: ICTC,
  IEEE, 2021, pp. 416--420.

\bibitem{jerbi2021quantum}
S.~Jerbi, L.~M. Trenkwalder, H.~P. Nautrup, H.~J. Briegel, V.~Dunjko, Quantum
  enhancements for deep reinforcement learning in large spaces, PRX Quantum
  2~(1) (2021) 010328.

\bibitem{airsim}
S.~Shah, D.~Dey, C.~Lovett, A.~Kapoor, Airsim: High-fidelity visual and
  physical simulation for autonomous vehicles, CoRR abs/1705.05065.

\bibitem{carla}
A.~Dosovitskiy, G.~Ros, F.~Codevilla, A.~M. L{\'{o}}pez, V.~Koltun, {CARLA}: An
  open urban driving simulator, CoRR abs/1711.03938.

\end{thebibliography}

\vspace{-0.3cm}

\end{document}